# An Information Theoretic Study for Noisy Compressed Sensing With Joint Sparsity Model–2

Sangjun Park, Nam Yul Yu and Heung-No Lee*, *Senior Member, IEEE*

*Abstract*—In this paper, we study a support set reconstruction problem in which the signals of interest are jointly sparse with a common support set, and sampled by joint sparsity model-2 (JSM-2) in the presence of noise. Using mathematical tools, we develop upper and lower bounds on the failure probability of support set reconstruction in terms of the sparsity, the ambient dimension, the minimum signal to noise ratio, the number of measurement vectors and the number of measurements. These bounds can be used to provide a guideline to determine the system parameters in various applications of compressed sensing with noisy JSM-2. Based on the bounds, we develop necessary and sufficient conditions for reliable support set reconstruction. We interpret these conditions to give theoretical explanations about the benefits enabled by joint sparsity structure in noisy JSM-2. We compare our sufficient condition with the existing result of noisy multiple measurement vectors model (MMV). As a result, we show that noisy JSM-2 may require less number of measurements than noisy MMV for reliable support set reconstruction.

*Index Terms*—compressed sensing, support set reconstruction, joint sparsity model–2, joint sparsity structure.

## I. INTRODUCTION

Conventionally, signals sensed from sensors such as microphones and imaging devices are sampled, following the Shannon and Nyquist sampling theory [1], at a rate higher than twice the maximum frequency for signal reconstruction. Since the number of samples decided by this theory is often large in volume, these samples go through a compression stage before being stored. Therefore, taking the large samples, where most of them will be thrown away in this stage, is inefficient. Compressed sensing (CS) [2] aims to remove this inefficiency.

The theory of CS states that any compressible signals that is sparsely representable in a certain basis can be compressively sampled and reconstructed from what we thought is incomplete information. Let $\mathbf{x} \in \mathbb{R}^N$ be a $K$ sparse vector with a support set $\mathcal{I} := \{i \mid x(i) \neq 0\}$ whose indices indicates the positions of its nonzero coefficients. This sparse vector can be compressively sampled by *single measurement vector model* (*SMV*)

$$\mathbf{y} = \mathbf{F}\mathbf{x} + \mathbf{n} \tag{1}$$

where $\mathbf{y} \in \mathbb{R}^M$ is a (noisy) measurement vector, $\mathbf{F} \in \mathbb{R}^{M \times N}$ is a sensing matrix and $\mathbf{n} \in \mathbb{R}^M$ is a noise vector whose elements are independent and identically distributed (i.i.d) Gaussian with zero mean and $\sigma^2$ variance. If the support set of $\mathbf{x}$ is known, (1) is simplified to

$$\mathbf{y} = \mathbf{F}_\mathcal{I}\mathbf{x}_\mathcal{I} + \mathbf{n}$$

where $\mathbf{F}_\mathcal{I} \in \mathbb{R}^{M \times K}$ is a submatrix formed by the columns of $\mathbf{F}$ indexed by $\mathcal{I}$, and $\mathbf{x}_\mathcal{I} \in \mathbb{R}^K$ is a subvector formed by the elements of $\mathbf{x}$ indexed by $\mathcal{I}$. We estimate $\mathbf{x}$ by solving the least square problem. Thus, estimating $\mathbf{x}$ is equal to solving a support set reconstruction problem, which has been extensively studied in the literature.

### A. Information-theoretic Works for CS with SMV

The works of [3] – [8] studied the support set reconstruction problem from an information-theoretic perspective. For reliable support set reconstruction, sufficient and necessary conditions were established in linear and sublinear sparsity regimes, i.e., $\lim_{N \to \infty} \frac{K}{N} \in (0, 1/2)$ and $\lim_{N \to \infty} \frac{K}{N} = 0$, respectively.

In support set reconstruction, Wainwright [3] used the union bound to establish a sufficient condition on $M$ for a maximum likelihood (ML) decoder and used the Fano's inequality [9] to get a necessary condition on $M$. This ML decoder was analyzed by Fletcher *et al.* [4] to establish a necessary condition on $M$. Aeron *et al.* [5] used the Fano's inequality to make necessary conditions on both $M$ and $\sigma^2$, while used the union bound to get sufficient conditions on both $M$ and $\sigma^2$ for their sub-optimal decoder.

Manuscript received April 01, 2016. This work was supported by the National Research Foundation of Korea (NRF) grant funded by the South Korean government (NRF-2015R1A2A1A05001826). This paper was presented in part at the International Symposium on Information Theory (ISIT), Boston, USA, 2012.

Sangjun Park, Nam Yul Yu, and Heung-No Lee are with the School of Electrical Engineering and Computer Science, Gwangju Institute of Science and Technology (GIST), Gwangju, 61005 Korea, e-mail: {sjpark, nyyu, heungno}@gist.ac.kr. The asterisk * indicates the corresponding author.

Akcakaya and Tarokh [6] used the union and the large deviation bounds based on empirical entropies to get sufficient conditions on *M* for their joint typical decoder. They also used the converse of the channel coding theorem to get necessary conditions on *M*. Scarlett *et al.* [7] extended this decoder [6] by assuming that the distribution of the support set is given. For the uniform distribution case, their necessary and sufficient conditions are equivalent to those of [6]. But, they are better for the non-uniform distribution case. Scarlett and Cevher [8] liked the support set reconstruction with the problem of coding over a mixed channel where information spectrum methods were used to get necessary and sufficient conditions on *M*.

*B. Information-theoretic Works for CS with MMV and JSM-2*

CS has many applications in wireless sensor networks (WSN) [10] – [13], and multiple inputs and multiple outputs (MIMO) radar [14]. In the applications, the signals of interest are often *jointly K sparse*, meaning that they have the joint support set, which is referred to as *joint (common) sparsity structure*.

Assume that all $\mathbf{x}^s \in \mathbb{R}^N$, $s = 1, 2, \cdots, S$, are jointly *K* sparse with a joint (common) support set $\mathcal{I}$. There are two models to sample them. In the first model, each sparse vector is sampled by a common sensing matrix, i.e.,

$$\mathbf{y}^s = \mathbf{F}\mathbf{x}^s + \mathbf{n}^s \in \mathbb{R}^M, \quad s = 1, 2, \cdots S$$

which is termed as *multiple measurement vectors model* (*MMV*) [15] where $\mathbf{n}^s$ is a noise whose elements are i.i.d. Gaussian with zero mean and $\sigma^2$ variance. In the second model, each one is sampled by its own sensing matrix, i.e.,

$$\mathbf{y}^s = \mathbf{F}^s \mathbf{x}^s + \mathbf{n}^s, \quad s = 1, 2, \cdots S$$

which is termed as *joint sparsity model-2* (*JSM-2*) [10][11]. Similar to SMV, one estimates all the sparse vectors in both MMV and JSM-2 by solving the least square problems, once the joint support set is known.

The authors of [16] – [18] conducted information-theoretic researches to get conditions under which a joint support set of both MMV and JSM-2 is reconstructed with high probability. In noisy MMV, Tang and Nehorai [16] used hypothesis theory to get necessary and sufficient conditions on both *M* and *S*, and proved that the success probability of support set reconstruction increases with *S* if $M = \Omega\left(K \log \frac{N}{K}\right)$. Jin and Rao [17] exploited communication theory to establish necessary and sufficient conditions on *M* in noisy MMV, and showed the benefits enabled by the joint sparsity structure based on their conditions. A detailed comparison between the results of this paper and [17] will be presented in Section IV. Last, Duarte *et al.* [18] studied noiseless JSM-2 and made necessary and sufficient conditions on *M*. But, it is difficult to apply the conditions to JSM-2 in the presence of noise.

Meanwhile, the works of [10][19][20] presented conditions for reliable support set reconstruction by developing practical algorithms. In noiseless MMV, Blanchard and Davies [20] got the conditions for reliable reconstruction from the rank-aware orthogonal matching pursuit (OMP). In noisy MMV, Kim *et al.* [20] made compressive MUSIC, and presented its sufficient condition. In noiseless JSM-2, Baron *et al.* [10] made trivial pursuit (TP) and distributed compressed sensing–simultaneous OMP (DCS-SOMP). They theoretically proved that $M > K$ is sufficient for TP to estimate all the sparse vectors when *S* goes to infinity. Besides, they empirically reported that $M > K$ is sufficient for DCS-SOMP to reliably reconstruct the support set when *S* goes to infinity.

To the best of our knowledge, no information-theoretic study has been published to get necessary and sufficient conditions for reliable support set reconstruction in noisy JSM-2. Besides, no one has provided these conditions from practical recovery algorithms for CS with noisy JSM-2.

*C. Motivation of This Paper*

CS with noisy JSM-2 has been applied in many applications, and the benefits enabled by the joint sparsity structure have been empirically reported. In WSN, Caione *et al.* [12] used the joint sparsity structure to reduce the number of transmitted bits per sensor, and reported that each sensor could reduce its transmission cost. In magnetic resonance imaging (MRI), Wu *et al.* [21] modeled multiple diffusion tensor images (DTI) as jointly sparse vectors. They exploited the joint sparsity structure to reduce the number of samples per DTI while the reconstruction quality was kept. By using the joint sparsity structure, they also empirically reported that the reconstruction quality of each DTI could be improved for a fixed number of samples per DTI.

To theoretically explain the above empirical benefits enabled by the joint sparsity structure, theoretical tools are required to measure the performance of CS with noisy JSM-2. Such tools can be used as a guideline to determine the system parameters in various applications of CS with noisy JSM-2. For example, if the number of samples of each DTI is fixed in MRI [21], theoretical tools may help us to determine the number of DTIs to achieve a given reconstruction quality. Thus, the first motivation of this paper is to provide theoretical tools by establishing sufficient and necessary conditions for reliable support set reconstruction.

Next, let us assume that *S* jointly *K* sparse vectors are given by $\mathbf{X} := \begin{bmatrix} \mathbf{x}^1 & \mathbf{x}^2 & \cdots & \mathbf{x}^S \end{bmatrix}$. In noiseless MMV and JSM-2, the measurement vectors $\mathbf{Y}_{MMV}$ and $\mathbf{Y}_{JSM-2}$ are then presented by

$$\mathbf{Y}_{MMV} := \begin{bmatrix} \mathbf{F}\mathbf{x}^1 & \mathbf{F}\mathbf{x}^2 & \cdots & \mathbf{F}\mathbf{x}^S \end{bmatrix} = \mathbf{F}\mathbf{X},$$

and

$$\mathbf{Y}_{JSM-2} := \begin{bmatrix} \mathbf{F}^1\mathbf{x}^1 & \mathbf{F}^2\mathbf{x}^2 & \cdots & \mathbf{F}^S\mathbf{x}^S \end{bmatrix}$$

where the sensing matrices $\mathbf{F}, \mathbf{F}^1, \cdots, \mathbf{F}^S$ have i.i.d. Gaussian elements with zero mean and unit variance. As $\mathbf{F}^1, \cdots, \mathbf{F}^S$ are independent, $\mathbf{Y}_{JSM-2}$ has uncorrelated elements. In contrast, $\mathbf{Y}_{MMV}$ has correlated elements because they are taken from the common sensing matrix $\mathbf{F}$. In addition, it is readily checked that $rank(\mathbf{Y}_{MMV}) \le rank(\mathbf{X}) \le \min(S,K)$ while $rank(\mathbf{Y}_{JSM-2}) = S$. When $S > K$, we have $\mathrm{rank}(\mathbf{Y}_{JSM-2}) > \mathrm{rank}(\mathbf{Y}_{MMV})$, which implies that more reliable support set reconstruction can be expected in noiseless JSM-2, if more measurement vectors than the sparsity are available. The second motivation is to verify this intuition in the presence of noise, by comparing our results with the existing ones in noisy MMV [17].

*D. Contributions of This Paper*

The contributions of this paper are summarized. First, we derive upper and lower bounds on the failure probability of support set reconstruction from Lemmas 1 and 2, by exploiting mathematical techniques such as the Fano's inequality [9] and the Chernoff bound [22]. The lower and upper bounds are used to measure the performance of CS with noisy JSM-2.

Second, we develop necessary and sufficient conditions for reliable support set reconstruction. In the linear sparsity regime, Theorem 1 states that

$$M = \Omega\left(K + \frac{K}{S}\right)$$

suffices to achieve reliable support set reconstruction with the failure probability approaching to zero. In the sublinear sparsity regime, Theorem 2 states that

$$M = \Omega\left(K + \frac{K}{S}\log\frac{N}{K}\right) \quad (2)$$

suffices to achieve reliable support set reconstruction with the failure probability approaching to zero. For finite $N$ and $K$, Theorem 4 states that

$$M = O\left(\frac{K}{S \log K}\log\frac{N}{K}\right)$$

is necessary for reliable support set reconstruction with the failure probability approaching to zero. The necessary and sufficient conditions can be used as a guideline to determine the system parameters of application of CS with noisy JSM-2. For fixed $N$ and $K$, Theorem 3 and Corollary 3 indicate that the sufficient condition is $M \ge K + 1$ for large $S$, which states that the fundamental limit of $M = K + 1$ is achieved, if infinitely many measurement vectors are taken. From the sufficient conditions, we provide theoretical explanations of the benefits enabled by the joint sparsity structure, which confirm the empirical results of the applications of CS with noisy JSM-2 [12][21]. Finally, we compare the sufficient condition (2) with the known one (48) for noisy MMV [17]. As a result, we show that if $S \ge K$, noisy JSM-2 may require less number of measurements for reliable support set reconstruction than noisy MMV under low-noise level scenario, which confirms the superiority of JSM-2.

## II. NOTATIONS, SYSTEM MODEL AND A JOINT TYPICAL DECODER

*A. Notations*

In this paper, $\mathbb{P}, \mathbb{E}$ and $\mathbb{V}$ denote the probability, expectation and (co)variance, respectively. A small (capital) bold letter $\mathbf{f}\,(\mathbf{F})$ is a vector (matrix). A subvector (submatrix) formed by the elements (columns) of $\mathbf{f}\,(\mathbf{F})$ indexed by a set $\mathcal{I}$ is $\mathbf{f}_\mathcal{I}\,(\mathbf{F}_\mathcal{I})$. The $i$th eigenvalue and the trace of $\mathbf{F}$ are denoted by $\lambda_i(\mathbf{F})$ and $\mathrm{tr}[\mathbf{F}]$, respectively. The $n$th derivative of a function $f(x)$ with respect to $x$ is $f^n(x)$. For given sets $\mathcal{I}$ and $\mathcal{J}$, the relative complement of $\mathcal{J}$ in $\mathcal{I}$ is denoted as $\mathcal{J} \setminus \mathcal{I}$. An orthogonal projection matrix of $\mathbf{F}$ is

$$\mathbf{Q}(\mathbf{F}) := \mathbf{I}_M - \mathbf{F}\left(\mathbf{F}^T\mathbf{F}\right)^{-1}\mathbf{F}^T$$

where $\mathbf{F}^T$ and $\mathbf{F}^{-1}$ are the transpose and the inverse of $\mathbf{F}$, respectively.

*B. System Model – JSM-2*

We let $\mathbf{x}^1, \mathbf{x}^2, \cdots, \mathbf{x}^S$ be jointly $K$ sparse vectors with a joint (common) support set $\mathcal{I}$ belonging to

$$\mathcal{S} := \{\mathcal{J} | \mathcal{J} \subset \{1, 2, \cdots, N\}, |\mathcal{J}| = K\}.$$

Each one is sampled by its own sensing matrix, i.e.,

$$\mathbf{y}^s = \mathbf{F}^s \mathbf{x}^s + \mathbf{n}^s \quad s = 1, 2, \cdots S \qquad (3)$$

where all the sensing matrices have i.i.d. Gaussian elements with zero mean and unit variance, and all the noise vectors have i.i.d. Gaussian elements with zero mean and $\sigma^2$ variance. We assume that all the noise vectors and all the sensing matrices are mutually independent, and let $x_{\min}$ be the smallest nonzero magnitude of all the sparse vectors. Then,

$$x_{\min}^2 = \min_{\mathcal{J} \in \mathcal{S} \setminus \mathcal{I}} \underbrace{\min_{s \in \{1,2,\cdots,S\}} \left\| \mathbf{x}_{\mathcal{I} \setminus \mathcal{J}}^s \right\|_2^2}_{=: x_{\min,\mathcal{J}}^2} \qquad (4)$$

is the minimum residual energy for any incorrect support set $\mathcal{J}$. We then define the minimum signal to noise ratio (SNR$_{\min}$) by

$$\text{SNR}_{\min} := x_{\min}^2 / \sigma^2.$$

*C. Joint Typical Decoder Analysis for Noisy JSM*

We extend Akcakaya and Tarokh [6]'s decoder for noisy JSM-2. It takes all the measurement vectors as its input, and yields a support set decision as its output

$$d : \{\forall_s (\mathbf{y}^s, \mathbf{F}^s)\} \mapsto \hat{\mathcal{I}} \in \mathcal{S}, \quad s = 1, 2, \cdots, S.$$

Its decision rules are given in Definition 1.

***Definition 1***: All the measurement vectors $\{\mathbf{y}^1, \mathbf{y}^2, \cdots, \mathbf{y}^S\}$ and a set $\mathcal{J} \in \mathcal{S}$ are $\delta$ *jointly typical* if $\forall_s \text{rank}(\mathbf{F}_\mathcal{J}^s) = K$ and

$$\left| \left( \sum_{s=1}^S \left\| \mathbf{Q}(\mathbf{F}_\mathcal{J}^s) \mathbf{y}^s \right\|_2^2 \right) - S(M-K)\sigma^2 \right| < SM\delta.$$

Define the failure events that the joint typical decoder fails to reconstruct the correct support set. First,

$$\mathcal{E}_\mathcal{I}^c := \left\{ \left| \left( \sum_{s=1}^S \left\| \mathbf{Q}(\mathbf{F}_\mathcal{I}^s) \mathbf{y}^s \right\|_2^2 \right) - S(M-K)\sigma^2 \right| \geq SM\delta \right\} \qquad (5)$$

implies that the correct support set is not $\delta$ jointly typical with all the measurement vectors. Second, for any $\mathcal{J} \in \mathcal{S} \setminus \mathcal{I}$,

$$\mathcal{E}_\mathcal{J} := \left\{ \left| \left( \sum_{s=1}^S \left\| \mathbf{Q}(\mathbf{F}_\mathcal{J}^s) \mathbf{y}^s \right\|_2^2 \right) - S(M-K)\sigma^2 \right| < SM\delta \right\} \qquad (6)$$

implies that an incorrect support set is $\delta$ jointly typical with all the measurement vectors. Based on these failure events, the failure probability and its upper bound are given by

$$\begin{aligned} p_{err} &:= \mathbb{P}\{\hat{\mathcal{I}} \neq \mathcal{I} | \mathbf{x}^1, \cdots, \mathbf{x}^S\} \\ &= \mathbb{P}\left\{ \mathcal{E}_\mathcal{I}^c \bigcup_{\mathcal{J} \in \mathcal{S} \setminus \mathcal{I}} \mathcal{E}_\mathcal{J} \right\} \\ &\leq \mathbb{P}\{\mathcal{E}_\mathcal{I}^c\} + \sum_{\mathcal{J} \in \mathcal{S} \setminus \mathcal{I}} \mathbb{P}\{\mathcal{E}_\mathcal{J}\} \end{aligned} \qquad (7)$$

where $\mathbb{P}\{\mathcal{E}_\mathcal{I}^c\}$ is taken with respect to all the noise vectors and $\mathbb{P}\{\mathcal{E}_\mathcal{J}\}$ is taken with respect to all the noise vectors and all the sensing matrices. In what follows, Lemmas 1 and 2 give upper bounds on the probabilities of the failure events by using the union and Chernoff [22] bounds.

*Lemma 1*: For any positive $\delta$, we have

$$\mathbb{P}\{\mathcal{E}_\mathcal{I}^c\} \leq 2\exp\left(-\frac{S(M-K)}{2}d_1\right)(1+d_1)^{\frac{S(M-K)}{2}} \tag{8}$$
$$=: 2p(d_1)$$

where

$$d_1 := \frac{M\delta}{(M-K)\sigma^2} > 0. \tag{9}$$

*Proof*: The proof is given in Appendix A.

*Lemma 2*: Let $\mathcal{J} \in \mathcal{S} \setminus \mathcal{I}$ and a matrix $\mathbf{R}_\mathcal{J}$ be

$$\mathbf{R}_\mathcal{J} = \begin{bmatrix} \alpha_{\mathcal{J},1}\mathbf{I}_{M-K} & & \\ & \ddots & \\ & & \alpha_{\mathcal{J},S}\mathbf{I}_{M-K} \end{bmatrix} \tag{10}$$

where

$$\alpha_{\mathcal{J},s} := \sigma^2 + \left\|\mathbf{x}_{\mathcal{I}\setminus\mathcal{J}}^s\right\|_2^2 > 0. \tag{11}$$

Consider any positive $\delta$ such that

$$0 < \delta < (1-K/M)(\lambda_{\min}(\mathbf{R}_\mathcal{J}) - \sigma^2)$$

where $\lambda_{\min}(\mathbf{R}_\mathcal{J})$ is the smallest eigenvalue of $\mathbf{R}_\mathcal{J}$. Then,

$$\mathbb{P}\{\mathcal{E}_\mathcal{J}\} \leq \exp\left(-\frac{S(M-K)}{2}\left(d_{2,\lambda_{\min}(\mathbf{R}_\mathcal{J})} - 1\right)\right) d_{2,\lambda_{\min}(\mathbf{R}_\mathcal{J})}^{\frac{S(M-K)}{2}}$$
$$=: p\left(d_{2,\lambda_{\min}(\mathbf{R}_\mathcal{J})}\right) \tag{12}$$
$$\leq p(d_{2,\alpha^*})$$

where

$$d_{2,\lambda_{\min}(\mathbf{R}_\mathcal{J})} := \frac{(M-K)\sigma^2 + M\delta}{(M-K)\lambda_{\min}(\mathbf{R}_\mathcal{J})} \in (0,1) \tag{13}$$

and

$$\alpha^* := \sigma^2 + x_{\min}^2. \tag{14}$$

*Proof*: The proof is given in Appendix A.

As will be stated in the proof of Lemma 2, the matrix in (10) is the covariance matrix of a multivariate Gaussian vector $\mathbf{b}$ in (60). Then for any incorrect support set, its smallest eigenvalue can be easily computed and lower bounded by

$$\lambda_{\min}(\mathbf{R}_\mathcal{J}) = \min_{s \in \{1,2,\cdots,S\}} \alpha_{\mathcal{J},s} = \sigma^2 + x_{\min,\mathcal{J}}^2 \geq \alpha^* \tag{15}$$

where $x_{\min,\mathcal{J}}^2$ is defined in (4).

Combining (7), (8) and (12) yields a final upper bound

$$p_{err} \leq 2p(d_1) + \sum_{\mathcal{J} \in \mathcal{S}\setminus\mathcal{I}} p\left(d_{2,\lambda_{\min}(\mathbf{R}_\mathcal{J})}\right)$$
$$\leq 2p(d_1) + \binom{N}{K}p(d_{2,\alpha^*}).$$

### III. MAIN RESULTS

As the main contribution of this paper, this section presents sufficient and necessary conditions on *M* for reliable support set reconstruction in noisy JSM-2. We then interpret the conditions to show the benefits enabled by the joint sparsity structure.

#### A. Sufficient Conditions on M

**Theorem 1**: For any $\rho > 1$, we let $\delta = \rho^{-1}(1 - K/M)x_{\min}^2$. If the number of measurements satisfies

$$M > K + \upsilon_1 \frac{K}{S} \tag{16}$$

then the failure probability $p_{err}$ defined in (7) converges to zero in the linear sparsity regime, where

$$\upsilon_1 = -\frac{2\left(1 - \log \frac{K}{N}\right)}{\log\left(1 - \frac{1 - \rho^{-1}}{1 + \text{SNR}_{\min}^{-1}}\right) + \frac{1 - \rho^{-1}}{1 + \text{SNR}_{\min}^{-1}}} > 0.$$

**Proof**: In the linear sparsity regime, *K* goes to infinity as *N* goes to infinity. Then, let $M = cK$ where $c > 1$. From (8),

$$\log \mathbb{P}\{\mathcal{E}_\mathcal{I}^c\} \le 2^{-1} SK(c-1)\underbrace{\left(\log(1 + d_1) - d_1\right)}_{=:A} + \log 2$$

where $A < 0$ due to (9). Thus,

$$\lim_{N \to \infty} \mathbb{P}\{\mathcal{E}_\mathcal{I}^c\} \le \lim_{K \to \infty} \exp\left(2^{-1} SK(c-1)A + \log 2\right) = 0$$

implying that the probability that the correct support set is not $\delta$ jointly typical with all the measurement vectors vanishes.

Next, from (12),

$$\begin{aligned}\log \sum_{\mathcal{J} \in \mathcal{S} \setminus \mathcal{I}} \mathbb{P}\{\mathcal{E}_\mathcal{J}\} &\le \log\left(\binom{N}{K} p(d_{2,\alpha^*})\right) \\ &= \log\binom{N}{K} + 2^{-1} SK(c-1)\underbrace{\left(\log(1-t) + t\right)}_{=:\gamma} \\ &\le K\underbrace{\left(1 + \log \tfrac{N}{K} + 2^{-1} S(c_1 - 1)\gamma\right)}_{=:\eta}\end{aligned} \tag{17}$$

where the last inequality is due to

$$\binom{N}{K} \le \exp\left(K \log \frac{Ne}{K}\right). \tag{18}$$

In (17), $\gamma < 0$ for any *t* where

$$t = \frac{1 - \rho^{-1}}{1 + \text{SNR}_{\min}^{-1}} \in (0,1). \tag{19}$$

If $c > 1 + S^{-1}\upsilon_1$, then $\eta < 0$, which yields

$$\lim_{N \to \infty} \sum_{\mathcal{J} \in \mathcal{S} \setminus \mathcal{I}} \mathbb{P}\{\mathcal{E}_\mathcal{J}\} \le \lim_{K \to \infty} \exp(K\eta) = 0$$

implying that the probability that all incorrect support sets are $\delta$ jointly typical with all the measurement vectors vanishes. Thus the failure probability $p_{err}$ defined in (7) converges to zero if *M* satisfies (16), which completes the proof. ∎

**Theorem 2**: For any $\rho > 1$, we let $\delta = \rho^{-1}(1 - K/M)x_{\min}^2$. If the number of measurements satisfies

$$M > K + \upsilon_2 \frac{K}{S} \log \frac{N}{K} \tag{20}$$

then the failure probability $p_{err}$ defined in (7) converges to zero in the sublinear sparsity regime, where

$$\upsilon_2 = -\frac{2}{\log\left(1 - \frac{1-\rho^{-1}}{1+\text{SNR}_{min}^{-1}}\right) + \frac{1-\rho^{-1}}{1+\text{SNR}_{min}^{-1}}} > 0.$$

*Proof*: Let $M = K + cK \log \frac{N}{K}$ wher $c > 1$. From (8),

$$\log \mathbb{P}\{\mathcal{E}_{\mathcal{I}}^c\} \leq 2^{-1} ScK \log \frac{N}{K} \underbrace{\left(\log(1+d_1) - d_1\right)}_{=:A} + \log 2$$

where $A < 0$ due to (9). Thus,

$$\lim_{N \to \infty} \mathbb{P}\{\mathcal{E}_{\mathcal{I}}^c\} \leq \lim_{N \to \infty} \exp\left(2^{-1} ScKA \log \frac{N}{K} + \log 2\right) = 0$$

implying that the probability that the correct support set is not $\delta$ jointly typical with all the measurement vectors vanishes.

Next, from (12),

$$\log \sum_{\mathcal{J} \in \mathcal{S} \setminus \mathcal{I}} \mathbb{P}\{\mathcal{E}_{\mathcal{J}}\} \leq \log\left(\binom{N}{K} p(d_{2,\alpha^*})\right)$$

$$= \log\binom{N}{K} + 2^{-1} ScK \underbrace{\left(\log(1-t) + t\right)}_{=:\gamma} \log \frac{N}{K}$$

$$\leq K \underbrace{\left(1 + 2^{-1} Sc\gamma\right)}_{=:\eta} \log \frac{N}{K} + K$$

where the last inequality is due to the bound in (18) and $\gamma < 0$ for any $t$ in (19). If $c > S^{-1}\upsilon_2$, then $\eta < 0$, which yields

$$\lim_{N \to \infty} \sum_{\mathcal{J} \in \mathcal{S} \setminus \mathcal{I}} \mathbb{P}\{\mathcal{E}_{\mathcal{J}}\} \leq \lim_{N \to \infty} \exp\left(K\eta \log \frac{N}{K} + K\right) = 0$$

implying that the probability that all incorrect support sets are $\delta$ jointly typical with all the measurement vectors vanishes. Thus, the failure probability $p_{err}$ defined in (7) converges to zero if $M$ satisfies (20), which completes the proof. ∎

From the sufficient conditions in both the theorems, we see an inverse relation between $M$ and $S$, due to the joint sparsity structure. This relation implies that taking more measurement vectors $S$ reduces the number of measurements $M$ required for reliable support set reconstruction. We then use it to explain the empirical results of Caione *et al.* [12] and Wu *et al.* [21]. Due to the usage of the joint sparsity structure, in [12], the authors reported that the number of transmitted bits per sensor could be inversely reduced by the number of sensors, which results in that the transmission cost of each sensor could be saved. The result can be confirmed by our inverse relation by considering $S$ and $M$ as the number of sensors and the number of transmitted bits per sensor, respectively. In [21], $S$ and $M$ are considered as the number of DTIs and the number of samples of each DTI, respectively. Again, it was observed from [21] that the joint sparsity structure enabled the number of samples of each DTI to be inversely reduced by the number of DTIs, which reduces the acquisition time of each DTI. This results also can be confirmed by our inverse relation.

Next, we consider a case where infinitely many measurement vectors are taken for fixed $N$ and $K$. The sufficient conditions of Theorems 1 and 2 cannot be used for this case because they are established when $N$ goes to infinity. In what follows, we present Theorem 3 to give a sufficient condition for fixed $N$ and $K$ with infinitely many measurement vectors.

*Theorem 3*: For any $\rho > 1$, we let $\delta = \rho^{-1}(1 - K/M)x_{min}^2$. If the number of measurements satisfies $M \geq K + 1$, the failure probability $p_{err}$ defined in (7) converges to zero as taking infinitely many measurement vectors, i.e., $S \to \infty$.

*Proof*: From Lemma 1,

$$\mathbb{P}\{\mathcal{E}_{\mathcal{I}}^c\} \leq 2\left(\underbrace{\exp\left(-\frac{M-K}{2}d_1\right)(1+d_1)^{\frac{M-K}{2}}}_{=:\mu_{\mathcal{I}}}\right)^S. \tag{21}$$

If $M \geq K + 1$, we have

$$\log \mu_{\mathcal{I}} = 2^{-1}(M-K)(\log(1+d_1) - d_1) < 0 \tag{22}$$

due to (9), which implies $\mu_{\mathcal{I}} < 1$. From Lemma 2,

$$\mathbb{P}\{\mathcal{E}_{\mathcal{J}}\} \leq \left(\underbrace{\exp\left(-\frac{M-K}{2}(d_{2,\alpha^*}-1)\right)d_{2,\alpha^*}^{\frac{M-K}{2}}}_{=:\mu_{\mathcal{J}}}\right)^S. \tag{23}$$

Similarly, if $M \geq K + 1$, we have

$$\log \mu_{\mathcal{J}} = 2^{-1}(M-K)(\log(1-t) + t) < 0 \tag{24}$$

due to (19), which implies $\mu_{\mathcal{J}} < 1$. Thus, we conclude

$$\lim_{S \to \infty} p_{err} \leq 2\lim_{S \to \infty} \mu_{\mathcal{I}}^S + \binom{N}{K}\lim_{S \to \infty} \mu_{\mathcal{J}}^S = 0$$

for $M \geq K + 1$ which completes the proof. ∎

In noiseless JSM-2, Baron *et al.* [10] showed that with $M > K$ measurements of each measurement vector, DCS-SOMP could reconstruct the support set when $S$ goes to infinity. This result is confirmed by Theorem 3.

### B. Discussions on Sufficient Conditions

We now study how $\text{SNR}_{\min}$ affects the sufficient conditions of Theorems 1 and 2. The aim is to reveal a relation among $S$, $M$ and $\text{SNR}_{\min}$ for reliable support set reconstruction.

***Corollary 1***: For any $\rho > 1$, we let $\delta = \rho^{-1}(1-K/M)x_{\min}^2$. The sufficient conditions of Theorems 1 and 2 are rewritten as

$$M > K + 4\left(\frac{\sqrt[-1]{S} + (\sqrt{S} \times \text{SNR}_{\min})^{-1}}{1-\rho^{-1}}\right)^2 K \log \frac{N}{K} \tag{25}$$

in the sublinear sparsity regime, and

$$M > K + 4K\left(\frac{\sqrt[-1]{S} + (\sqrt{S} \times \text{SNR}_{\min})^{-1}}{1-\rho^{-1}}\right)^2 \left(1 - \log \frac{K}{N}\right) \tag{26}$$

in the linear sparsity regime.

***Proof***: From the inequality $\log(1+x) \leq \frac{2x}{2+x}$ for $x \in (-1, 0]$,

$$\upsilon_2 = -\frac{2}{\log(1-t)+t} < \frac{4-2t}{t^2} < \frac{4}{t^2} \tag{27}$$

where $t$ is defined in (19). Then,

$$\frac{\upsilon_2}{S} = -\frac{2}{\log(1-t)+t} \times \frac{1}{S} \leq \frac{4}{St^2}. \tag{28}$$

From (19),

$$\sqrt{S}t = \frac{1-\rho^{-1}}{\sqrt[-1]{S}+\left(\sqrt{S}\times\text{SNR}_{\min}\right)^{-1}}. \tag{29}$$

Combining (20), (28) and (29) leads to (25). This approach is used to get (26) using the following equality

$$\upsilon_1 = \upsilon_2\left(1-\log\tfrac{N}{K}\right) \tag{30}$$

which completes the proof. ∎

Corollary 1 suggests that for a fixed $M$, reliable support set reconstruction is possible by taking more measurement vectors $S$ though $\text{SNR}_{\min}$ is low. Namely, we observe a noise reduction effect, which shows that using the joint sparsity structure leads to increasing $\text{SNR}_{\min}$ or reducing $\sigma^2$ by a factor of $S$. This effect can explain that the reconstruction quality of the DTIs could be improved, as empirically reported in [21].

We then improve our noise reduction effect by considering that $\text{SNR}_{\min}$ is larger than a certain value.

***Corollary 2***: For any $\rho>1$, we let $\delta=\rho^{-1}(1-K/M)x_{\min}^2$ and $\alpha\in(0,1-\rho^{-1})$ be a constant. If

$$\text{SNR}_{\min} \geq \frac{\alpha}{1-\rho^{-1}-\alpha}, \tag{31}$$

the sufficient conditions of Theorems 1 and 2 are rewritten as

$$M > K + \frac{\left(S^{-1}+\left(S\times\text{SNR}_{\min}\right)^{-1}\right)(4-2\alpha)}{(1-\rho^{-1})\alpha}K\log\frac{N}{K} \tag{32}$$

in the sublinear sparsity regime, and

$$M > K + \frac{\left(S^{-1}+\left(S\times\text{SNR}_{\min}\right)^{-1}\right)(4-2\alpha)}{(1-\rho^{-1})\alpha}K\left(1-\log\frac{K}{N}\right) \tag{33}$$

in the linear sparsity regime.

***Proof***: From (27) with (31),

$$\upsilon_2 = -\frac{2}{\log(1-t)+t} < \frac{4-2t}{t^2} \leq \frac{2}{t}\left(\frac{2-\alpha}{\alpha}\right)$$

where $t$ is defined in (19). Then,

$$\frac{\upsilon_2}{S} = -\frac{2}{\log(1-t)+t}\times\frac{1}{S} < \frac{4-2\alpha}{St\alpha}. \tag{34}$$

From (19),

$$St = \frac{1-\rho^{-1}}{S^{-1}+\left(S\times\text{SNR}_{\min}\right)^{-1}}. \tag{35}$$

Combining (20), (34) and (35) leads to (32). This approach is used to get (33) using (30), which completes the proof. ∎

Remind that Theorem 3 shows that the fundamental limit of $M=K+1$ is achieved for reliable support set reconstruction when we take infinitely many measurement vectors. But, taking infinite measurement vectors is unrealistic. As an example, we cannot deploy infinitely many sensors in a restricted region for WSN. To give a bound on $S$ for achieving the fundamental limit, we presrent Corollary 3.

***Corollary 3***: For any $\rho>1$, we let $\delta=\rho^{-1}(K+1)^{-1}x_{\min}^2$ and the fundamental limit of $M=K+1$ be achieved. If the number of measurement vectors satisfies

$$S > \left(\log\left(\binom{N}{K}\right)+2\right)-\log\varepsilon\right) \times \max\left[\left|\frac{1}{\log\mu_{\mathcal{I}}}\right|, \left|\frac{1}{\log\mu_{\mathcal{J}}}\right|\right] \quad (36)$$

reliable support set reconstruction is possible, i.e., $p_{err} < \varepsilon$ for sufficiently small $\varepsilon \in (0,1)$, where $\mu_{\mathcal{I}}$ and $\mu_{\mathcal{J}}$ are defined in (21) and (23), respectively. The lower bound on $S$ is decreasing with respect to $\text{SNR}_{\min}$.

*Proof*: We assume that $\mu_{\mathcal{I}} \geq \mu_{\mathcal{J}}$ and

$$p_{err} \leq \mathbb{P}\{\mathcal{E}_{\mathcal{I}}^c\} + \sum_{\mathcal{J} \in \mathcal{S}\setminus\mathcal{I}} \mathbb{P}\{\mathcal{E}_{\mathcal{J}}\} \leq \left(\binom{N}{K}+2\right)\mu_{\mathcal{I}}^S < \varepsilon. \quad (37)$$

Then, if the number of measurement vectors satisfies

$$S > \frac{\log\varepsilon - \log\left(\binom{N}{K}+2\right)}{\log\mu_{\mathcal{I}}}, \quad (38)$$

(37) is achieved for small $\varepsilon$, and hence, reliable support set reconstruction is possible. If $\mu_{\mathcal{I}} < \mu_{\mathcal{J}}$, we obtain inequalities similar to (37) and (38) by replacing $\mu_{\mathcal{I}}$ by $\mu_{\mathcal{J}}$, where

$$S > \frac{\log\varepsilon - \log\left(\binom{N}{K}+2\right)}{\log\mu_{\mathcal{J}}}. \quad (39)$$

Combining (38) and (39) yields (36).

Next, a simple computation yields that for any $d_1$ in (9),

$$\frac{\partial \log \mu_{\mathcal{I}}}{\partial d_1} = -\frac{d_1}{2(1+d_1)} < 0$$

where $\log\mu_{\mathcal{I}}$ is given in (22). From (9), we see $d_1 \propto \text{SNR}_{\min}$ that leads to $\log\mu_{\mathcal{I}} \propto \text{SNR}_{\min}^{-1}$. Also, for any $t$ in (19),

$$\frac{\partial \log \mu_{\mathcal{J}}}{\partial t} = -\frac{t}{2(1-t)} < 0$$

where $\log\mu_{\mathcal{J}}$ is given in (24). From (19), we see $t \propto \text{SNR}_{\min}$ that leads to that $\log\mu_{\mathcal{J}} \propto \text{SNR}_{\min}^{-1}$. Hence, the lower bound on $S$ in (36) turns out to be a decreasing function with respect to $\text{SNR}_{\min}$, which completes the proof. ∎

Now, we assume that $\text{SNR}_{\min}$ goes to infinity. From (22) and (24), it is easy to see that

$$\lim_{\text{SNR}_{\min} \to \infty} \log\mu_{\mathcal{I}} = -\infty,$$

$$\lim_{\text{SNR}_{\min} \to \infty} \log\mu_{\mathcal{J}} = 2^{-1}\left(1 - \rho^{-1} - \log\rho\right).$$

Hence, (36) is simplified to

$$S > \left(\log\left(\binom{N}{K}+2\right)-\log\varepsilon\right) \times \left|2\left(1-\rho^{-1}-\log\rho\right)^{-1}\right|. \quad (40)$$

Note that $N$, $K$ and $\varepsilon$ are fixed. Thus for large $\rho$, we have

$$\left|1 - \rho^{-1} - \log\rho\right| \gg 2\left(\log\left(\binom{N}{K}+2\right)-\log\varepsilon\right), \quad (41)$$

which leads to $S > 0$. Thus if $\text{SNR}_{\min}$ goes to infinity and $\rho$ satisfies (41), the fundamental limit of $M = K + 1$ is achieved regardless of the number of measurement vectors.

*C. Necessary Condition on M*

We give a necessary condition that any decoder must satisfy for reliable support set reconstruction in noisy JSM-2. Unlike our sufficient conditions of Theorems 1 and 2, this necessary condition is established for finite $N$ and $K$.

We begin by transforming (3) into

$$\begin{bmatrix} \mathbf{y}^1 \\ \vdots \\ \mathbf{y}^S \end{bmatrix}_{=:\mathbf{y} \in \mathbb{R}^{SM}} = \underbrace{\begin{bmatrix} \mathbf{F}^1 & & \\ & \ddots & \\ & & \mathbf{F}^S \end{bmatrix}}_{=:\mathbf{F} \in \mathbb{R}^{SM \times SN}} \begin{bmatrix} \mathbf{x}^1 \\ \vdots \\ \mathbf{x}^S \end{bmatrix}_{=:\mathbf{x} \in \mathbb{R}^{SN}} + \begin{bmatrix} \mathbf{n}^1 \\ \vdots \\ \mathbf{n}^S \end{bmatrix}_{=:\mathbf{n} \in \mathbb{R}^{SM}} \tag{42}$$

where $\mathbf{x}$ is an $SK$ sparse vector and it belongs to an infinite set

$$\mathcal{X}_{x_{\min}} := \left\{ \mathbf{x} \in \mathbb{R}^{SN} \big| |x(i)| \ge x_{\min}, \forall i \in \mathcal{I}, |\mathcal{I}| = SK \right\}$$

where $x(i)$ is the $i$th element of $\mathbf{x}$ and $\mathcal{I}$ is the support set of $\mathbf{x}$. Thanks to the joint sparsity structure, the number of possible support set is $\binom{N}{K}$. Then, define the failure probability

$$p_{err} := \mathbb{E}_{\mathbf{F}} \sup_{\mathbf{x} \in \mathcal{X}_{x_{\min}}} \mathbb{P}\left\{ \hat{\mathcal{I}} \ne \mathcal{I} \big| \mathbf{x}, \mathbf{F} \right\} \tag{43}$$

where $\hat{\mathcal{I}}$ is an estimate of the support set based on $\mathbf{y}$ and $\mathbf{F}$ in (42). Lemma *III*-3 of [5] yields

$$\sup_{\mathbf{x} \in \mathcal{X}_{x_{\min}}} \mathbb{P}\left\{ \hat{\mathcal{I}} \ne \mathcal{I} \big| \mathbf{x}, \mathbf{F} \right\} \ge \min_{\hat{\mathbf{x}} \in \mathcal{X}_{\{x_{\min}\}}} \max_{\mathbf{x} \in \mathcal{X}_{\{x_{\min}\}}} \mathbb{P}\left\{ \hat{\mathbf{x}} \ne \mathbf{x} \big| \mathbf{x}, \mathbf{F} \right\} \tag{44}$$

where $\hat{\mathbf{x}}$ is an estimate for $\mathbf{x}$ based on $\mathbf{y}$ and $\mathbf{F}$ in (42) and

$$\mathcal{X}_{\{x_{\min}\}} := \left\{ \mathbf{x} \in \mathbb{R}^{SN} \big| x(i) = x_{\min}, \forall i \in \mathcal{I}, |\mathcal{I}| = SK \right\}$$

which is a finite set. Assume that $\mathbf{x}$ is uniformly distributed over this finite set. Applying the Fano's inequality [9] to (44) yields

$$\max_{\mathbf{x} \in \mathcal{X}_{\{x_{\min}\}}} \mathbb{P}\left\{ \hat{\mathbf{x}} \ne \mathbf{x} \big| \mathbf{x}, \mathbf{F} \right\} \ge \mathbb{P}\left\{ \hat{\mathbf{x}} \ne \mathbf{x} \big| \mathbf{F} \right\}$$
$$\ge 1 - \frac{\mathbb{I}(\mathbf{x}; \mathbf{y} | \mathbf{F}) + \log 2}{\log\left(\left|\mathcal{X}_{\{x_{\min}\}}\right| - 1\right)} \tag{45}$$

where $\mathbf{x}$ and $\hat{\mathbf{x}}$ belong to the finite set $\mathcal{X}_{\{x_{\min}\}}$ and $\mathbb{I}(\mathbf{x}; \mathbf{y})$ is the mutual information between $\mathbf{x}$ and $\mathbf{y}$. In what follows, we get a necessary condition on $M$ to ensure that the lower bound in (45) is bounded away from zero.

***Theorem 4***: In (42), if the number of measurements satisfies

$$M < \frac{2K \log \frac{N}{K} - 2 \log 2}{S \log(1 + K \times \mathrm{SNR}_{\min})} \tag{46}$$

then the failure probability $p_{err}$ defined in (43) is bounded away from zero.

***Proof***: The mutual information in (45) is bounded by

$$\mathbb{I}(\mathbf{x}; \mathbf{y} | \mathbf{F}) = h(\mathbf{y} | \mathbf{F}) - h(\mathbf{y} | \mathbf{x}, \mathbf{F}) \le h(\mathbf{y}) - h(\mathbf{n})$$
$$\le \sum_{i=1}^{SM} h(y_i) - h(\mathbf{n})$$
$$\le 2^{-1} SM \left( \log\left(2\pi e \left(K x_{\min}^2 + \sigma^2\right)\right) - \log\left(2\pi e \sigma^2\right) \right)$$
$$= 2^{-1} SM \log(1 + K \times \mathrm{SNR}_{\min})$$

where $h(\mathbf{x})$ is the differential entropy of $\mathbf{x}$, and $h(\mathbf{x} | \mathbf{y})$ is the conditional entropy of $\mathbf{x}$ given $\mathbf{y}$. The last inequality is due to that the Gaussian distribution maximizes the differential entropy. The denominator in (45) is bounded by

$$\log\left(\left|\mathcal{X}_{\{x_{\min}\}}\right|-1\right) = \log\left(\binom{N}{K}-1\right) > K\log\frac{N}{K}$$

for sufficiently large $N$. Then,

$$\begin{aligned} p_{err} &= \mathbb{E}_{\mathbf{F}} \sup_{\mathbf{x} \in \mathcal{X}_{\{x_{\min}\}}} \mathbb{P}\left\{\hat{\mathcal{I}} \neq \mathcal{I} \middle| \mathbf{x}, \mathbf{F}\right\} \\ &\geq \mathbb{E}_{\mathbf{F}} \min_{\hat{\mathbf{x}} \in \mathcal{X}_{\{x_{\min}\}}} \max_{\mathbf{x} \in \mathcal{X}_{\{x_{\min}\}}} \mathbb{P}\left\{\hat{\mathbf{x}} \neq \mathbf{x} \middle| \mathbf{x}, \mathbf{F}\right\} \\ &> 1 - \frac{2^{-1}SM\log\left(1+K\times\text{SNR}_{\min}\right)+\log 2}{K\log\frac{N}{K}}. \end{aligned} \quad (47)$$

From (47), the failure probability is bounded away from zero by zero if (46) is satisfied, which completes the proof. ∎

## IV. RELATIONS TO EXISTING INFORMATION THEORETIC RESULTS

### A. Relation to Noisy MMV model [17]

Jin and Rao [17] exploited the Chernoff bound to get a tight sufficient condition on $M$ for reliable support set reconstruction for noisy MMV in the sublinear sparsity regime. Owing to the complicated form of their sufficient condition, they could not clearly show the benefits enabled by the joint sparsity structure. Thus, they simplified their condition under scenarios such as *i*) a low noise level scenario and *ii*) an $S$ identical sparse vectors scenario. We compare the sufficient condition (2) with these conditions.

First, in a low noise level scenario, the sufficient condition [17] for noisy MMV is

$$M = \Omega\left(\frac{K\log N}{\min(K,S)}\right). \quad (48)$$

If $S < K$, the sufficient conditions (2) and (48) have the same order, which implies no significant performance gap in support set reconstruction between noisy JSM-2 and MMV. But if $S \geq K$, (48) is $M = \Omega(\log N)$, while (2) is $M = \Omega(S^{-1}K\log N)$, which demonstrates that noisy JSM-2 can be superior to noisy MMV for $S \geq K$ with respect to the number of measurements required for reliable support set reconstruction. This comparison result supports the intuition presented in Section I-C, where more reliable support set reconstruction could be expected in noiseless JSM-2, due to the linear independency of the measurement vectors. Moreover, it validates the intuition even in the presence of noise.

Second, in noisy MMV with an $S$ identical sparse vectors scenario, the sufficient condition of [17] is

$$M = \Omega\left(\frac{K\log N}{\log\left(1+S\|\mathbf{x}\|_2^2/\sigma^2\right)}\right). \quad (49)$$

From (49), we observe that $\sigma^2$ is reduced by a factor of $S$. But, the noise reduction effect of noisy MMV requires a restriction that all sparse vectors are identical, which is hardly achieved in practice. In contrast, the noise reduction effect of noisy JSM-2 does not require the restriction, as shown in Corollaries 1 and 2.

### B. Relation to Noisy SMV model [6]

Akcakaya and Tarokh [6] used the joint typical decoder to establish the sufficient conditions on $M$ for reliable support set reconstruction in noisy SMV. They exploited the exponential inequalities [23] to obtain upper bounds on the sum of weighted chi-square random variables. In this subsection, we aim to show that the approaches developed in this paper is superior to the use of the exponential inequalities. Thus we use the exponential inequalities to generalized their bounds for noisy JSM-2. Then, we establish Propositions 1 and 2 to prove that the generalized bounds are worse than the bounds of Lemmas 1 and 2.

*Proposition 1*: For any positive $\delta$, we have

$$\mathbb{P}\left\{\mathcal{E}_{\mathcal{I}}^c\right\} \leq 2p(d_1) \leq 2p_{1,\exp}$$

where both $p(d_1)$ and $d_1$ are given in Lemma 1, and

$$p_{1,\exp} := \exp\left(-\frac{S\delta^2}{4\sigma^4}\frac{M^2}{M-K+2\delta M/\sigma^2}\right). \tag{50}$$

*Proof*: The proof is given in Appendix B.

***Proposition 2***: For any $\mathcal{J} \in \mathcal{S} \setminus \mathcal{I}$ and any $\delta > 0$ such that

$$0 < \delta < (1-K/M)x_{\min,\mathcal{J}}^2, \tag{51}$$

we have

$$\mathbb{P}\{\mathcal{E}_\mathcal{J}\} \le p\left(d_{2,\lambda_{\min}(\mathbf{R}_\mathcal{J})}\right) \le p_{2,\mathcal{J},\exp}$$

where $p\left(d_{2,\lambda_{\min}(\mathbf{R}_\mathcal{J})}\right)$ and $d_{2,\lambda_{\min}(\mathbf{R}_\mathcal{J})}$ are given in Lemma 2 and

$$p_{2,\mathcal{J},\exp} := \exp\left(-\frac{S^2(M-K)}{4\sum_{s=1}^S \alpha_{\mathcal{J},s}^2}\left(x_{\min,\mathcal{J}}^2 - \frac{M\delta}{M-K}\right)^2\right) \tag{52}$$

and $\alpha_{\mathcal{J},s}$ is defined in (11).

*Proof*: The proof is given in Appendix B.

If $S = 1$, we can see that $p_{1,\exp}$ and $p_{2,\mathcal{J},\exp}$ are equivalent to the bounds of Akcakaya and Tarokh [6]. Propositions 1 and 2 state that the bounds on the failure probability of Lemmas 1 and 2 are tighter than the bounds of [6] for noisy SMV.

## V. Conclusions

We have studied a support set reconstruction problem for CS with noisy JSM-2. The union and the Chernoff bounds have been used to get an upper bound on the failure probability of support set reconstruction, and the Fano's inequality has been used to get a lower bound on this failure probability. Since we have obtained the upper bound by analyzing an exhaustive search decoder, the bound is used to measure the performance of CS with noisy JSM-2. We then have developed the necessary and sufficient conditions in terms of the sparsity $K$, the ambient dimension $N$, the number of measurements $M$, the number of measurement vectors $S$, and the minimum signal to noise ratio $\text{SNR}_{\min}$. These conditions can be used as a guideline to decide the system parameters in various applications of CS with noisy JSM-2.

Interpreting the conditions provides theoretical explanations to the benefits enabled by the joint sparsity structure in noisy JSM-2.
  i. From the sufficient conditions of Theorems 1 and 2, we have observed an inverse relation between $M$ and $S$. Thanks to the inverse relation, we can take less samples per each measurement vector for reliable support set reconstruction by taking more measurement vectors.
  ii. From the sufficient conditions of Corollaries 1 and 2, we have observed a noise reduction effect, which shows that the usage of the joint sparsity structure results in increasing $\text{SNR}_{\min}$ or reducing $\sigma^2$ by a factor of $S$. As a result, support set reconstruction can be robust as taking more measurement vectors.
  iii. From Theorem 3 and Corollary 3, we have shown that the fundamental limit can be achieved for fixed $N$ and $K$ when $S$ is sufficiently large.

The above theoretical explanations confirmed the empirical benefits enabled by the joint sparsity structure, as shown in the applications of CS with noisy JSM-2 [12][21].

We have compared our sufficient conditions for noisy JSM-2 with other existing results [17] for noisy MMV. For a low-level noise scenario with $S \ge K$, we have shown that the number of measurements for reliable support set reconstruction of noisy JSM-2 is less than that for noisy MMV. Also, we have observed that noisy JSM-2 enjoys a noise reduction effect for arbitrary sparse vectors, while this noise reduction exists in noisy MMV only if all the sparse vectors are identical.

Recently, assuming that a signal of interest lies in a union of subspaces, a problem of reconstructing such signals sampled in noisy SMV has been studied. Wimalajeewa *et al.* [24] further reduced the number of measurements when the subspace has *block sparsity structure*. In the future work, we will assume that all sparse vectors have joint and block sparsity structures, and are sampled by noisy JSM-2. Under this assumption, we will study the problem of support set reconstruction to get necessary and sufficient conditions for understanding the performance of noisy JSM-2 when both the structures are jointly exploited.

## APPENDIX A
### PROOF OF LEMMAS 1 AND 2

First of all, we give the Scharf's theorem [25] to compute the moment generating function of a quadratic random variable. We then make Lemmas 3 and 4 to give the moment generating functions of the random variables of $\mathcal{E}_\mathcal{I}^c$ and $\mathcal{E}_\mathcal{J}$ that will be used in the proofs of Lemmas 1 and 2, respectively.

***Scharf's theorem [Page 64, [25]]***: Let $\mathbf{b} \in \mathbb{R}^N$ be a multivariate Gaussian vector with mean $\mathbf{m}$ and covariance $\mathbf{R}$. Then a random variable $Q \triangleq (\mathbf{b}-\mathbf{m})^T(\mathbf{b}-\mathbf{m})$ is quadratic with $\mathbb{E}[Q] = \text{tr}[\mathbf{R}]$, $\mathbb{V}[Q] = 2\text{tr}[\mathbf{R}^T\mathbf{R}]$ and for any $t$

$$\mathbb{E}[\exp(tQ)] = \prod_{i=1}^{N}(1-2t\lambda_i(\mathbf{R}))^{-1/2}.$$

***Lemma 3***: In (5), define a quadratic random variable

$$Z_\mathcal{I} := \sum_{s=1}^{S} \left\| \mathbf{Q}(\mathbf{F}_\mathcal{I}^s)\mathbf{y}^s \right\|_2^2 \Big/ \sigma^2. \tag{53}$$

Then, $\mathbb{E}[Z_\mathcal{I}] = S(M-K)$, $\mathbb{V}[Z_\mathcal{I}] = 2S(M-K)$ and for any $0 < t < 0.5$,

$$\mathbb{E}[\exp(tZ_\mathcal{I})] = (1-2t)^{-S(M-K)/2}. \tag{54}$$

***Proof***: The orthogonal projection matrix is decomposed as

$$\mathbf{Q}(\mathbf{F}_\mathcal{I}^s) = \mathbf{U}_\mathcal{I}^s \mathbf{D}^s (\mathbf{U}_\mathcal{I}^s)^T$$

where $\mathbf{D}^s$ is a diagonal matrix and $\mathbf{U}_\mathcal{I}^s$ is a unitary matrix. Each diagonal matrix has $M-K$ nonzero elements of one, and each unitary matrix is independent of each noise vector. Then,

$$Z_\mathcal{I} = \sum_{s=1}^{S} \left\| \mathbf{Q}(\mathbf{F}_\mathcal{I}^s)\mathbf{y}^s \right\|_2^2 \Big/ \sigma^2 = \sum_{s=1}^{S} \left\| \mathbf{Q}(\mathbf{F}_\mathcal{I}^s)\mathbf{n}^s \right\|_2^2 \Big/ \sigma^2$$

$$= \sum_{s=1}^{S} \left\| \mathbf{D}^s \underbrace{(\mathbf{U}_\mathcal{I}^s)^T \mathbf{n}^s \big/\sigma^2}_{=:\mathbf{w}^s} \right\|_2^2 = \sum_{s=1}^{S} \left\| \mathbf{D}^s \mathbf{w}^s \right\|_2^2 \tag{55}$$

where $\mathbf{w}^s$ is a multivariate Gaussian vector with mean $\mathbf{0}_M$ and covariance $\mathbf{I}_M$. Let the nonzero elements of all $\mathbf{D}^s$ be on the first $M-K$ diagonals. Then,

$$Z_\mathcal{I} = \sum_{s=1}^{S}\left\|\mathbf{D}^s\mathbf{w}^s\right\|_2^2 = \sum_{s=1}^{S}\sum_{i=1}^{M-K}\left|w^s(i)\right|^2$$

$$= \sum_{s=1}^{S}(\mathbf{w}_\mathcal{P}^s)^T\mathbf{w}_\mathcal{P}^s = \mathbf{w}^T\mathbf{w} \tag{56}$$

which is quadratic, where

$$\mathbf{w}_\mathcal{P}^s = \begin{bmatrix} w^s(1) & w^s(2) & \cdots & w^s(M-K) \end{bmatrix}^T$$

and

$$\mathbf{w} = \begin{bmatrix} (\mathbf{w}_\mathcal{P}^1)^T & (\mathbf{w}_\mathcal{P}^2)^T & \cdots & (\mathbf{w}_\mathcal{P}^S)^T \end{bmatrix}^T. \tag{57}$$

In (55), $\mathbf{w}^s$ is determined by $\mathbf{U}_\mathcal{I}^s$ and $\mathbf{n}^s$. Since the elements of $\mathbf{U}_\mathcal{I}^s$ and $\mathbf{n}^s$ are independent, $\mathbf{w}^i$ and $\mathbf{w}^j$ are mutually independent for any $1 \leq i \neq j \leq S$. The covariance matrix of $\mathbf{w}$ is an identity matrix. Thus, applying the Scharf's theorem to $Z_\mathcal{I}$ completes the proof. ∎

***Lemma 4***: In (6), for any $\mathcal{J} \in \mathcal{S} \setminus \mathcal{I}$, define a quadratic random variable

$$Z_\mathcal{J} := \sum_{s=1}^{S}\left\|\mathbf{Q}(\mathbf{F}_\mathcal{J}^s)\mathbf{y}^s\right\|_2^2. \tag{58}$$

Then, $\mathbb{E}[Z_\mathcal{J}] = \text{tr}[\mathbf{R}_\mathcal{J}]$, $\mathbb{V}[Z_\mathcal{J}] = 2\text{tr}[\mathbf{R}_\mathcal{J}^T \mathbf{R}_\mathcal{J}]$ and for any $t$,

$$\mathbb{E}[\exp(tZ_\mathcal{J})] = \prod_{i=1}^{S(M-K)} (1 - 2t\lambda_i(\mathbf{R}_\mathcal{J}))^{-1/2},$$

where $\mathbf{R}_\mathcal{J}$ is given in (10).

*Proof*: Similar to the proof of Lemma 3,

$$\mathbf{Q}(\mathbf{F}_\mathcal{J}^s) = \mathbf{U}_\mathcal{J}^s \mathbf{D}^s (\mathbf{U}_\mathcal{J}^s)^T$$

where $\mathbf{D}^s$ is a diagonal matrix and $\mathbf{U}_\mathcal{J}^s$ is a unitary matrix. Each $\mathbf{D}^s$ has $M - K$ nonzero elements of one, and each $\mathbf{U}_\mathcal{I}^s$ is independent of $\{\mathbf{f}_u^s : u \in \mathcal{I} \setminus \mathcal{J}\}$ and $\mathbf{n}^s$. Then,

$$\begin{aligned} Z_\mathcal{J} &= \sum_{s=1}^{S} \|\mathbf{Q}(\mathbf{F}_\mathcal{J}^s) \mathbf{y}^s\|_2^2 = \sum_{s=1}^{S} \|\mathbf{Q}(\mathbf{F}_\mathcal{J}^s) \mathbf{c}^s\|_2^2 \\ &= \sum_{s=1}^{S} \left\|\mathbf{D}^s \underbrace{(\mathbf{U}_\mathcal{J}^s)^T \mathbf{c}^s}_{=:\mathbf{b}^s}\right\|_2^2 = \sum_{s=1}^{S} \|\mathbf{D}^s \mathbf{b}^s\|_2^2 \end{aligned} \quad (59)$$

where $\mathbf{b}^s$ is a multivariate Gaussian vector with mean $\mathbf{0}_M$ and

$$\mathbb{V}[\mathbf{b}^s] = \left(\sigma^2 + \|\mathbf{x}_{\mathcal{I}\setminus\mathcal{J}}^s\|_2^2\right) \mathbf{I}_M$$

and $\mathbf{c}^s = \mathbf{n}^s + \sum_{u \in \mathcal{I} \setminus \mathcal{J}} \mathbf{f}_u^s x^s(u)$. Let the nonzero elements of all $\mathbf{D}^s$ be on the first $M - K$ diagonals. Then,

$$\begin{aligned} Z_\mathcal{J} &= \sum_{s=1}^{S} \|\mathbf{D}^s \mathbf{b}^s\|_2^2 = \sum_{s=1}^{S} \sum_{i=1}^{M-K} |b^s(i)|^2 \\ &= \sum_{s=1}^{S} (\mathbf{b}_\mathcal{P}^s)^T \mathbf{b}_\mathcal{P}^s = \mathbf{b}^T \mathbf{b} \end{aligned} \quad (60)$$

which is quadratic, where

$$\mathbf{b}_\mathcal{P}^s = \begin{bmatrix} b^s(1) & b^s(2) & \cdots & b^s(M-K) \end{bmatrix}^T$$

and

$$\mathbf{b} = \begin{bmatrix} (\mathbf{b}_\mathcal{P}^1)^T & (\mathbf{b}_\mathcal{P}^2)^T & \cdots & (\mathbf{b}_\mathcal{P}^S)^T \end{bmatrix}^T.$$

In (59), $\mathbf{b}^s$ is determined by $\mathbf{U}_\mathcal{J}^s$, $\mathbf{n}^s$ and $\{\mathbf{f}_u^s : u \in \mathcal{I} \setminus \mathcal{J}\}$. Since the elements of $\mathbf{U}_\mathcal{J}^s$, $\mathbf{n}^s$ and $\{\mathbf{f}_u^s : u \in \mathcal{I} \setminus \mathcal{J}\}$ are independent, $\mathbf{b}^i$ and $\mathbf{b}^j$ are mutually independent for any $1 \leq i \neq j \leq S$. The covariance matrix of $\mathbf{b}$ is diagonal as shown in (10). Thus, applying the Scharf's theorem to $Z_\mathcal{J}$ completes the proof. ∎

Next, we give proofs of Lemmas 1 and 2.

*Proof of Lemma 1*: From (5), we have

$$\mathbb{P}\{\mathcal{E}_\mathcal{I}^c\} = \mathbb{P}\{Z_\mathcal{I} \leq W_1\} + \mathbb{P}\{Z_\mathcal{I} \geq W_2\} \quad (61)$$

where

$$W_i = S(M-K) + (-1)^i SM\delta/\sigma^2, \quad i = 1, 2.$$

Applying the Chernoff bound [22] to (61) yields

$$\mathbb{P}\{\mathcal{E}_\mathcal{I}^c\} \leq \sum_{i=1}^{2} \exp(-t_i W_i) \mathbb{E}\left[\exp(t_i Z_\mathcal{I})\right]$$
$$= \sum_{i=1}^{2} \underbrace{\exp(-t_i W_i)(1-2t_i)^{-S(M-K)/2}}_{=:f(t_i;W_i)} \quad (62)$$

where the equality is from Lemma 3, $t_1 < 0$ and $t_2 \in (0, \tfrac{1}{2})$. As each $f(t_i; W_i)$ is convex, $t_i = t_i^*$ at $f^{(1)}(t_i; W_i) = 0$ yields the minimizer of $f(t_i; W_i)$, where

$$t_i^* = 2^{-1}\left(1 - W_i^{-1} S(M-K)\right), \quad i = 1, 2.$$

Thus, $f(t_i; W_i) \geq f(t_i^*; W_i)$ for each $i$. If $W_1 \leq 0$, it is clear that $\mathbb{P}\{Z_\mathcal{I} \leq W_1\} = 0$ because $Z_\mathcal{I}$ is quadratic. Thus,

$$\mathbb{P}\{\mathcal{E}_\mathcal{I}^c\} = \mathbb{P}\{Z_\mathcal{I} \geq W_2\} \leq f(t_2^*; W_2) = p(d_1) \quad (63)$$

where $p(d_1)$ and $d_1$ are defined in (8) and (9), respectively. If $W_1 > 0$ then $f(t_1^*; W_1) \leq f(t_2^*; W_2)$ because

$$\log f(t_1^*; W_1) - \log f(t_2^*; W_2)$$
$$= S(M-K)\left[d_1 + 2\log(1-d_1) - 2\log(1+d_1)\right] < 0.$$

Thus,

$$\mathbb{P}\{\mathcal{E}_\mathcal{I}^c\} = f(t_1^*; W_1) + f(t_2^*; W_2) \leq 2f(t_2^*; W_2) = 2p(d_1). \quad (64)$$

Finally, combining (63) and (64) leads to (8). ∎

*Proof of Lemma 2*: From (6), we have

$$\mathbb{P}\{\mathcal{E}_\mathcal{J}\} = \mathbb{P}\{Z_\mathcal{J} < W_1\} - \mathbb{P}\{Z_\mathcal{J} < W_2\} \leq \mathbb{P}\{Z_\mathcal{J} < W_1\} \quad (65)$$

where

$$W_i = S(M-K)\sigma^2 - (-1)^i SM\delta, \quad i = 1, 2. \quad (66)$$

Applying the Chernoff bound [22] to (65) yields for $t < 0$,

$$\mathbb{P}\{\mathcal{E}_\mathcal{J}\} \leq \exp(-tW_1) \mathbb{E}\left[\exp(tZ_\mathcal{J})\right]$$
$$= \exp(-tW_1) \prod_{i=1}^{S(M-K)} (1 - 2t\lambda_i(\mathbf{R}_\mathcal{J}))^{-1/2} \quad (67)$$
$$=: p_{\mathcal{J}, JSM-2}(t)$$

where the equality is from Lemma 4. Remind that all the eigenvalues are positive. Thus for any $t < 0$,

$$p_{\mathcal{J}, JSM-2}(t) \leq \exp(-tW_1)(1 - 2t\lambda_{\min}(\mathbf{R}_\mathcal{J}))^{-S(M-K)/2}$$
$$=: f(t; W_1). \quad (68)$$

We then define a function $h(t) := \log f(t; W_1)$. Then,

$$h^{(2)}(t) = 2S(M-K)\lambda_{\min}^2(\mathbf{R}_\mathcal{J})(1 - 2t\lambda_{\min}(\mathbf{R}_\mathcal{J}))^{-2} > 0$$

which implies that $h$ is convex with respect to $t$. It leads to that $f$ in (68) is logarithmically convex. Thus $t = t^*$ at $f^{(1)}(t; W_1) = 0$ yields the minimizer of $f(t; W_1)$ where

$$t^* = 2^{-1}\left(\lambda_{\min}^{-1}(\mathbf{R}_\mathcal{J}) - W_1^{-1} S(M-K)\right) < 0.$$

Substituting $t^*$ in (68) together with (67) yields

$$\mathbb{P}\{\mathcal{E}_{\mathcal{J}}\} \leq f(t^*; W_1) = p\left(d_{2,\lambda_{\min}(\mathbf{R}_{\mathcal{J}})}\right). \tag{69}$$

where $p\left(d_{2,\lambda_{\min}(\mathbf{R}_{\mathcal{J}})}\right)$ and $d_{2,\lambda_{\min}(\mathbf{R}_{\mathcal{J}})}$ are defined in (12) and (13), respectively.

We prove $p\left(d_{2,\lambda_{\min}(\mathbf{R}_{\mathcal{J}})}\right) \leq p(d_{2,\alpha^*})$. Let $\beta = 2^{-1}S(M-K)$ and $x = d_{2,\lambda_{\min}(\mathbf{R}_{\mathcal{J}})}$ in the upper bound (69). Then, we have $p(x) = x^{\beta} \exp(-\beta(x-1))$, where

$$\frac{\partial p(x)}{\partial x} = \beta x^{\beta} \exp(-\beta(x-1))(x^{-1}-1) > 0 \tag{70}$$

and

$$\frac{\partial x}{\partial \lambda_{\min}(\mathbf{R}_{\mathcal{J}})} = -x < 0. \tag{71}$$

Due to (70) and (71),

$$\frac{\partial p(x)}{\partial \lambda_{\min}(\mathbf{R}_{\mathcal{J}})} = \frac{\partial p(x)}{\partial x} \frac{\partial x}{\partial \lambda_{\min}(\mathbf{R}_{\mathcal{J}})}$$
$$= -\beta x^{\beta-1} \exp(-\beta(x-1))(x^{-1}-1) < 0$$

which shows that the upper bound in (69) is decreasing with respect to $\lambda_{\min}(\mathbf{R}_{\mathcal{J}})$. For any $\mathcal{J} \in \mathcal{S} \setminus \mathcal{I}$, due to (15), we have

$$p\left(d_{2,\lambda_{\min}(\mathbf{R}_{\mathcal{J}})}\right) = p(x) \leq p(d_{2,\alpha^*})$$

which completes the proof. ∎

## APPENDIX B
## PROOF OF PROPOSITIONS 1 AND 2

First of all, we introduce the exponential inequalities [23], and use them in the proofs of Propositions 1 and 2.

*The exponential inequalities [23]*: Let $Y_i, i=1,2,\cdots,D$ be i.i.d. Gaussian variables with zero mean and unit variance. Then, let $\alpha_i, i=1,2\cdots,D$ be non-negative. We set

$$|\alpha|_{\infty} = \sup|\alpha_i|, \quad |\alpha|_2^2 = \sum_{i=1}^{D}\alpha_i^2$$

and let

$$Y = \sum_{i=1}^{D}\alpha_i(Y_i^2-1). \tag{72}$$

Then, the following inequalities hold for any positive $x$

$$\mathbb{P}\{Y \geq 2|\alpha|_2\sqrt{x} + 2|\alpha|_{\infty}x\} \leq \exp(-x) \tag{73}$$

$$\mathbb{P}\{Y \leq -2|\alpha|_2\sqrt{x}\} \leq \exp(-x). \tag{74}$$

*Proof of Proposition 1*: In the proof of Lemma 3, $Z_{\mathcal{I}}$ is represented by

$$Z_{\mathcal{I}} = \sum_{s=1}^{S}\sum_{i=1}^{M-K} w^s(i)^2$$

where $w^s(i)$ is Gaussian with zero mean and unit variance. Define a random variable $Y$ as

$$Y = Z_{\mathcal{I}} - S(M-K)$$
$$= \sum_{s=1}^{S} \sum_{i=1}^{M-K} \left( w^s(i)^2 - 1 \right).$$

which is of the form of (72). Then,

$$\mathbb{P}\{\mathcal{E}_{\mathcal{I}}^c\} = \underbrace{\mathbb{P}\{Y \leq -SM\delta/\sigma^2\}}_{=:A} + \underbrace{\mathbb{P}\{Y \geq SM\delta/\sigma^2\}}_{=:B}.$$

Combining $A$ with (74) gives

$$\mathbb{P}\{Y \leq -SM\delta/\sigma^2\} = \mathbb{P}\{Y \leq -2\sqrt{S(M-K)}x\}$$
$$\leq \underbrace{\exp\left(-\frac{SM^2\delta^2}{4(M-K)\sigma^4}\right)}_{=:C}$$

and combining $B$ with (73) gives

$$\mathbb{P}\{Y \geq SM\delta/\sigma^2\} = \mathbb{P}\{Y \geq 2\sqrt{S(M-K)}x + 2x\}$$
$$\leq p_{1,\exp}$$

where $p_{1,\exp}$ is defined in (50). It is readily seen that $p_{1,\exp} \geq C$, which leads to $\mathbb{P}\{\mathcal{E}_{\mathcal{I}}^c\} \leq 2p_{1,\exp}$.

Next, from (8) and (50),

$$\log p(d_1) = 2^{-1} S(M-K)\left(\log(1+d_1) - d_1\right)$$

and

$$\log p_{1,\exp} = -2^{-1} S(M-K) d_1^2 (2+4d_1)^{-1}$$

where $d_1 > 0$ is defined in (9). Then, we have

$$\log \frac{p(d_1)}{p_{1,\exp}} = \frac{S(M-K)}{2} \underbrace{\left(\log(1+d_1) - d_1 + d_1^2(2+4d_1)^{-1}\right)}_{=:g(d_1)}.$$

For any $d_1 > 0$, $\frac{\partial g(d_1)}{\partial d_1} = -d_1^2(2+3d_1)(1+d_1)^{-1}(1+2d_1)^{-2} < 0$ and $\max_{d_1 > 0} g(d_1) = 0$. Thus, we conclude $\log \frac{p(d_1)}{p_{1,\exp}} \leq 0$, which completes the proof. ∎

*Proof of Proposition 2*: In the proof of Lemma 4, $Z_{\mathcal{J}}$ is represented by

$$Z_{\mathcal{J}} = \sum_{s=1}^{S} \sum_{i=1}^{M-K} b^s(i)^2$$
$$= \sum_{s=1}^{S} \sum_{i=1}^{M-K} \alpha_{\mathcal{J},s} g^s(i)^2$$

where $\alpha_{\mathcal{J},s}$ is defined in (11) and $g^s(i)$ is Gaussian with zero mean and unit variance. Define a new random variable $Y$ as

$$Y = Z_{\mathcal{J}} - S(M-K)$$
$$= \sum_{s=1}^{S} \sum_{i=1}^{M-K} \alpha_{\mathcal{J},s} \left( g^s(i)^2 - 1 \right).$$

which is of the form of (72). Then, from (65)

$$\mathbb{P}\{\mathcal{E}_\mathcal{J}\} \leq \mathbb{P}\left\{Y < SM\delta - (M-K)\sum_{s=1}^{S}\left\|\mathbf{x}_{\mathcal{I}\setminus\mathcal{J}}^{s}\right\|_2^2\right\}$$

$$\leq \mathbb{P}\left\{Y < \underbrace{SM\delta - S(M-K)x_{\min,\mathcal{J}}^2}_{=:A}\right\} \quad (75)$$

$$\leq p_{2,\mathcal{J},\exp}$$

where $p_{2,\mathcal{J},\exp}$ is defined in (52), the last inequality is due to (74). Due to (51), $A$ is negative. Thus the exponential inequality of (74) gives the upper bound $p_{2,\mathcal{J},\exp}$.

Next, from (12) and (52),

$$\log p\left(d_{2,\lambda_{\min}(\mathbf{R}_\mathcal{J})}\right) = 2^{-1}S(M-K)(t+\log(1-t))$$

and

$$\log p_{2,\mathcal{J},\exp} \geq -\frac{S(M-K)}{4}\left(\frac{x_{\min,\mathcal{J}}^2 - \frac{M\delta}{M-K}}{x_{\min,\mathcal{J}}^2 + \sigma^2}\right)^2$$

$$= -4^{-1}S(M-K)t^2$$

where $t \in (0,1)$ is defined in (19) and the inequality is due to (15). Then,

$$\log \frac{p\left(d_{2,\lambda_{\min}(\mathbf{R}_\mathcal{J})}\right)}{p_{2,\mathcal{J},\exp}} \leq \frac{S(M-K)}{4}\underbrace{\left(t^2 + 2t + 2\log(1-t)\right)}_{=:g(t)}.$$

For any $t \in (0,1)$, $\frac{\partial g(t)}{\partial t} = -2t^2(1-t)^{-1} < 0$ and $\max_{t\in(0,1)} g(t) = 0$. Thus, we conclude $\log \frac{p\left(d_{2,\lambda_{\min}(\mathbf{R}_\mathcal{J})}\right)}{p_{2,\mathcal{J},\exp}} \leq 0$, which completes the proof. ∎